\documentclass[10pt,twocolumn]{article} 
\usepackage{simpleConference}
\usepackage{times}
\usepackage{graphicx}
\usepackage{amssymb}
\usepackage{url,hyperref}
\usepackage{balance}
\graphicspath{{fig/}}
\usepackage{algorithmic}
\usepackage{algorithm}
\usepackage{booktabs}
\usepackage{multicol}
\usepackage{subcaption}
\usepackage{acro}
\usepackage{tablefootnote}
\usepackage{siunitx}
\usepackage{xspace}
\usepackage{multirow}
\newcommand{\mr}[1]{\multirow{2}{*}}
\usepackage{cite}
\usepackage{amsmath,amssymb,amsfonts}
\usepackage{textcomp}
\usepackage{comment}
\usepackage{adjustbox}
\usepackage[capitalize]{cleveref}
\crefformat{equation}{(#2#1#3)}
\crefmultiformat{equation}{(#2#1#3)}{ and~(#2#1#3)}{, (#2#1#3)}{ and~(#2#1#3)}
\crefrangemultiformat{equation}{(#3#1#4)--(#5#2#6)}{,(#3#1#4)--(#5#2#6)}{,(#3#1#4)--(#5#2#6)}{,(#3#1#4)--(#5#2#6)}
\crefrangeformat{equation}{(#3#1#4)--(#5#2#6)}
\usepackage[scale=1.0]{inconsolata}
\usepackage{microtype}
\usepackage{footmisc}

\DeclareAcronym{asic}{ short = ASIC, long = application-specific integrated circuit, indefinite = an }
\DeclareAcronym{bn}{ short = BN, long = batch normalization }
\DeclareAcronym{bram}{ short = BRAM, long = BlockRAM }
\DeclareAcronym{cnn}{ short = CNN, long = convolutional neural network }
\DeclareAcronym{cpu}{ short = CPU, long = central processing unit }
\DeclareAcronym{ddr}{ short = DDR, long = double data rate }
\DeclareAcronym{dma}{ short = DMA, long = direct memory access }
\DeclareAcronym{dnn}{ short = DNN, long = deep neural network }
\DeclareAcronym{dpu}{ short = DPU, long = deep learning processor unit }
\DeclareAcronym{dsp}{ short = DSP, long = digital signal processor block }
\DeclareAcronym{fc}{ short = FC, long = fully connected, short-indefinite = an }
\DeclareAcronym{fifo}{ short = FIFO, long = first in first out }
\DeclareAcronym{fpga}{ short = FPGA, long = field-programmable gate array, short-indefinite = an }
\DeclareAcronym{fps}{ short = FPS, long = frame per second, short-indefinite = an, plural-form = frames per second }
\DeclareAcronym{gpu}{ short = GPU, long = graphical processing unit }
\DeclareAcronym{hbm}{ short = HBM, long = high bandwidth memory }
\DeclareAcronym{hls}{ short = HLS, long = high-level synthesis }
\DeclareAcronym{ilp}{ short = ILP, long = integer linear programming, indefinite = an }
\DeclareAcronym{ip}{ short = IP, long = intellectual property, indefinite = an }
\DeclareAcronym{lut}{ short = LUT, long = lookup table }
\DeclareAcronym{mac}{ short = MAC, long = multiply and accumulate }
\DeclareAcronym{madc}{ short = MADC, long = merge add to conv }
\DeclareAcronym{mcc}{ short = MCC, long = merge conv to conv }
\DeclareAcronym{ml}{ short = ML, long = machine learning }
\DeclareAcronym{mqc}{ short = MQC, long = merge quant to conv }
\DeclareAcronym{mqq}{ short = MQQ, long = merge quant to quant }
\DeclareAcronym{msc}{ short = MSC, long = merge skip to conv }
\DeclareAcronym{nn}{ short = NN, long = neural network, short-indefinite = an }
\DeclareAcronym{ocm}{ short = OCM, long = on-chip memory }
\DeclareAcronym{pe}{ short = PE, long = processing element }
\DeclareAcronym{rnn}{ short = RNN, long = recurrent neural network }
\DeclareAcronym{rom}{ short = ROM, long = read-only memory }
\DeclareAcronym{rtl}{ short = RTL, long = register-transfer level, short-indefinite = an }
\DeclareAcronym{sgd}{ short = SGD, long = stochastic gradient descent }
\DeclareAcronym{simd}{ short = SIMD, long = single instruction multiple data }
\DeclareAcronym{uram}{ short = URAM, long = UltraRAM }

\begin{document}

\title{Design and Optimization of Residual Neural Network Accelerators for Low-Power FPGAs Using High-Level Synthesis}

\author{ \\
\uppercase{Filippo Minnella},
\uppercase{Teodoro Urso}\\
\uppercase{Mihai T. Lazarescu}\\
\uppercase{Luciano Lavagno}
}

\maketitle

\begin{abstract}

Residual \aclp{nn} (ResNets) are widely used in computer vision tasks.
They enable the construction of deeper and more accurate models by mitigating the vanishing gradient problem.
Their main innovation is the \emph{residual block} which allows the output of one layer to bypass one or more intermediate layers and be added to the output of a later layer.
Their complex structure and the buffering required by the 
residual block makes them difficult to implement on 
resource-constrained platforms.
We present a novel design flow for implementing deep learning models for \acp{fpga} optimized for ResNets, using a strategy to reduce their buffering 
overhead to obtain a resource-efficient implementation of the 
residual layer.
The current implementations of residual networks suffer from diminished performance and heightened computational latency attributable to the way residual blocks are implemented.
Our \acl{hls} based flow encompasses a thorough set of design principles and optimization strategies, exploiting in novel ways standard techniques such as \emph{temporal reuse} and \emph{loop merging} to efficiently map ResNet models, and potentially other skip connection-based \ac{nn} architectures, into FPGA.
The models are quantized to 8-bit integers for both weights and 
activations, 16 bits for biases, and 32 bits for accumulations.
The experimental results are obtained on the CIFAR-10 dataset 
using ResNet8 and ResNet20 implemented with Xilinx FPGAs using 
\ac{hls} on the Ultra96-V2 and Kria KV260 boards.
Compared to the state-of-the-art on the Kria KV260 board, our ResNet20 implementation achieves 
\SI{2.88}{\times} speedup with \SI{0.5}{\percent} higher accuracy 
of \SI{91.3}{\percent}, while ResNet8 accuracy improves by 
\SI{2.8}{\percent} to \SI{88.7}{\percent}.
The throughputs of ResNet8 and ResNet20 are \SI{12971}{FPS} and 
\SI{3254}{FPS} on the Ultra96 board, and \SI{30153}{FPS} and 
\SI{7601}{FPS} on the Kria KV26, respectively.
They Pareto-dominate state-of-the-art solutions with respect to accuracy, throughput, and energy.
\end{abstract}

\acresetall
\section{Introduction}


\Acp{cnn} have consistently achieved state-of-the-art results in 
many tasks, including computer vision and speech recognition 
\cite{review0}. Their success is based on high accuracy and 
performance due to the improved computational intensity of 
convolutional layers compared to previous approaches, requiring 
less memory bandwidth than \acf{fc} layers \cite{CNN2015}.
The choice of hardware for implementing convolutional layers 
profoundly impacts their applicability. \Acp{cpu} are versatile and easy to program, but their architecture makes them relatively inefficient. \Acp{gpu} are designed to handle massive parallelism, allowing them to process multiple computations simultaneously. This aligns well with the inherently parallel nature of \acp{cnn} but their energy consumption is notably higher \cite{Dhilleswararao}.
\acp{asic} and \acp{fpga} offer different 
tradeoffs of cost and flexibility for algorithm acceleration 
\cite{acclerators}.
The latter are less performance and energy efficient due to their reprogrammability, but they have much lower design cost and can be more easily customized for a specific application.
%
%
\Acp{nn} optimized for embedded applications 
\cite{mobilenet,shuf} are designed to run efficiently on devices 
with limited processing power, memory, and energy. They can 
perform very well on small datasets, such as CIFAR-10 
\cite{cifar10} and MNIST \cite{deng2012mnist}, and are often used 
in real-time contexts where response timeliness and low latency 
are critical.

\begin{figure}
    \centering
    \includegraphics[width=0.5\columnwidth]{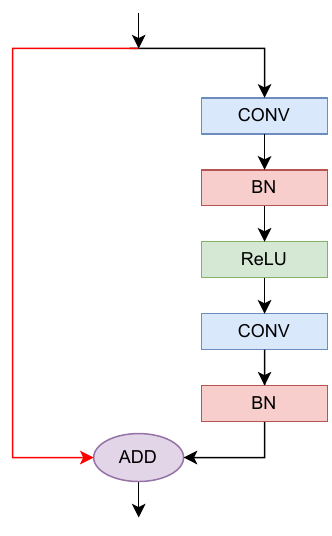}
    \caption{Basic residual block with a long branch with two 
    convolutional layers, and the skip connection (red branch) 
    that must store its input activations until output activation 
    is generated, requiring much memory.}
    \label{fig:resblock}
\end{figure}

Residual \aclp{nn} (ResNets) \cite{Resnet} use residual blocks 
(see \cref{fig:resblock}) to mitigate the vanishing gradient 
problem for deep networks through \emph{skip connections}. They 
allow intermediate feature maps to be reprocessed at different 
points in the network computation, increasing accuracy.
However, state-of-the-art implementations of \emph{skip connections} require significant 
on-chip buffering resources, which significantly reduce the benefits of 
streaming-based \ac{fpga} implementations. Thus, recent work has 
focused on optimizing and shrinking the residual structure of 
\acp{nn} \cite{NonResnet} and on finding quantization strategies 
that improve the efficiency of their hardware design \cite{9774574}.

Deep networks have many parameters and require extensive 
quantization to reduce their size to fit into the \ac{fpga} 
on-chip memory \cite{SA-BNN}. For this reason, widely used tools such as 
FINN \cite{FINN} focus on low-bit quantization (such as 1-bit 
\cite{BNN2020} or 2-bit \cite{choi2019accurate}) and make 
suboptimal use of resources for higher-bit quantization, as we will show later. However, 
low-bit quantizations degrade \ac{nn} accuracy and may not be 
suitable for accurate inference for complex problems 
\cite{UmurogluFGBLJV16}.
We propose an efficient residual block architecture for a concurrent dataflow \ac{fpga} 
implementation, in which each layer is a process and activations are streamed between  processes, with the following main contributions:
\begin{itemize}
    \item An optimized architecture for \acp{cnn} 
    that supports residual networks and minimizes buffering 
    resources, allowing on-chip storage of the parameters and activations using 8 bits, which have been shown to achieve good accuracy for our target \ac{nn} architectures \cite{LIANG2021370}.

    \item A custom \ac{hls} code generation flow from the Python 
    model to the \ac{fpga} bitstream using Vitis HLS 
    \cite{vitishls}. 
    \item The validation of the architecture and the 
    implementation flow using the ResNet8 and ResNet20 residual 
    \acp{nn} on CIFAR-10 targeting the Ultra96-V2 and Kria KV260 
    \ac{fpga} boards, to demonstrate the advantages of the 
    proposed solution.
\end{itemize}

The rest of the paper is organized as follows. \cref{sec:related 
work} presents the background and motivation for this work. 
\cref{sec:meth} discusses training and quantization, and 
describes the accelerator architecture with a special focus on 
skip connection management. \cref{sec:experiment} presents the 
experimental setup and discusses the results. \cref{sec:conc} 
concludes the paper.


\section{Related Work}
\label{sec:related work}

The field of FPGA-based acceleration for deep neural networks has gained significant attention due to its potential to achieve high-performance and energy-efficient inference. Several approaches and architectures have been proposed in the literature to address this challenge \cite{review1}.
In systolic array overlay-based architectures, each \ac{pe} is a \ac{simd} vector accumulation module, which receives activation inputs and weights in each cycle from the horizontally and vertically adjacent \acp{pe}.\\
Pipelined groups of \acp{pe} with short local communication and 
regular architectures can achieve high clock frequencies and 
efficient global data transfers \cite{Xuechao2017}. The overlay 
architecture \cite{Yu2020} performs the computation of the 
convolution layers in sequence over a systolic array. However, 
despite its flexibility, it has high latency due to frequent 
transfers between external memory [\acf{ddr} or \acf{hbm}] and 
on-chip memory.\\
An alternative approach is to implement a \emph{custom dataflow 
architecture where each layer is associated with a compute unit}. 
This structure can be pipelined, and activations and weights can 
be stored in \ac{ocm}, reducing latency and increasing 
throughput. The main limitation of this type of architecture is 
the number of \acp{dsp} and \acp{lut} required to implement the 
convolutional layers, as well as the size of on-chip buffers for 
weight storage \cite{fpgadataflowefficient}, while activations are streamed from layer to layer.
Since streaming tasks have well-defined pipelining rates with static data flows, a customized approach can lead to optimized processing, resulting in improved performance and resource saving.

Widely recognized as one of the leading frameworks for deploying \acp{dnn} on \acp{fpga}, Xilinx Vitis AI \cite{VitisAI} provides a comprehensive set of tools specifically designed for optimizing and deploying \acp{dnn} on Xilinx \acp{fpga}. With support for 
popular frameworks such as TensorFlow \cite{tensorflow}, PyTorch \cite{torch}, and Caffe \cite{caffe}, it incorporates various optimization techniques such as pruning, quantization, and kernel 
fusion to improve performance and reduce memory consumption.
The \acf{dpu} is the accelerator core used in Vitis AI and consists of several key modules, including a high-performance scheduler module, a hybrid computing array module, an instruction fetch unit module, and a global memory pool module \cite{DPUCZDX8G}. The \ac{dpu} is responsible for executing the microcode of the specified \ac{dnn} model, known as the 
\emph{xmodel}.
Vitis AI uses an overlay-based architecture where model weights and biases are stored in \ac{ddr} memory and cached in the on-chip weight buffer during inference. Input and output data of the \ac{pe} array are also cached in \ac{ocm}. This architecture scales very well for \ac{dnn}, but may have higher resource utilization and less performance compared to custom dataflow accelerators due to its general-purpose nature and the overhead associated with off-chip memory accesses.

Another widely used tool is FINN \cite{FINN}, an open source framework developed by Xilinx that allows the generation of highly optimized \acp{dnn} for \ac{fpga} acceleration, with emphasis on dataflow-style architectures. FINN uses \ac{hls} to convert trained \ac{dnn} models into hardware \ac{ip} blocks that can be easily integrated into \ac{fpga}-based systems.
While FINN offers significant customization capabilities, it is primarily designed for low-bitwidth quantization schemes, such as binarized networks. Achieving high performance with FINN often leads to lower accuracy and/or higher resources, particularly when using the 8-bit quantization that has been shown to be the best compromise between resources and accuracy.

\cite{Ushiroyama2022} evaluates the accuracy, power, throughput, and design time of three different \acp{cnn} implemented on \acp{fpga} and compares these metrics to their \ac{gpu} equivalents. A comparison was also made between a custom 
implementation of two \acp{dnn} using System Verilog and an implementation using the Xilinx tools FINN and Vitis AI \cite{Machura2022}. In addition, \cite{framework2023} reports a comparison between FINN and Vitis AI using a widely used set of 
ResNet model configurations.

We propose an optimized pipelined dataflow architecture tailored 
for better resource management. Our solution specifically targets 
residuals \acp{nn} and allows using more bits during quantization than e.g.~FINN, 
to improve the tradeoff between 
accuracy, throughput, and memory usage \cite{wu2020integer}. Its 
effectiveness is compared with Vitis AI, FINN, and custom 
resource-efficient approaches \cite{Zhang2022}, demonstrating its 
potential to efficiently implement complex residual \acp{nn} with 
state-of-the-art performance in terms of latency, energy, 
accuracy, and resource consumption.

\section{Methodology}
\label{sec:meth}

\begin{figure}
\centering
\includegraphics[width=0.3\columnwidth]{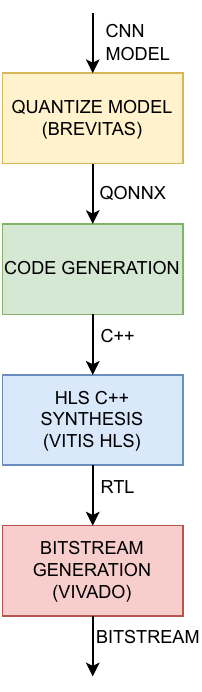}
\caption{Implementation flow}
\label{fig:flow}
\end{figure}

\cref{fig:flow} shows the high-level view of the flow that we use 
to generate C++ code from the quantized \ac{nn} model, which 
includes the following main steps:
\begin{itemize}
    \item Use Brevitas \cite{brevitas} for \ac{nn} quantization 
    and extract its graph in QONNX format, which provides an 
    easy-to-parse description of the network, including 
    information such as layer type, input and output quantization, and 
    layer connections (see \cref{sec:meth:quant});

    \item Generate the C++ of the top function that instantiates 
    all the layers of the network (see \cref{sec:meth:graph} and 
    \cref{sec:meth:net});

    \item Generate the \ac{rtl} code using Vitis HLS and a set of 
    model-independent synthesizable C++ libraries that we wrote 
    to implement the optimized layers (see \cref{sec:meth:conv});

    \item Import the \ac{rtl} code as a block into a Vivado 
    design and generate the bitstream for \ac{fpga} programming.

\end{itemize}

\subsection{Quantization}
\label{sec:meth:quant}

Quantization is done using the Brevitas framework 
\cite{brevitas}. \Acp{nn} are trained using PyTorch, and both the 
weights ($\mathit{bw}_\mathrm{w}$) and the activations 
($\mathit{bw}_\mathrm{x}$) are represented as 8-bit integers, 
because for a variety of applications this is a good compromise 
between memory usage and accuracy (\cite{LIANG2021370}), while the 
biases are represented as 16-bit integers 
($\mathit{bs}_\mathrm{b}$) for the same reason.

\Ac{nn} training uses floating-point calculations and models 
quantization by clamping and rounding. Back-propagation 
uses dequantized inputs and weights to improve convergence and 
accuracy, while loss evaluation uses quantization to match the 
results of the hardware implementation. Inference on the 
\ac{fpga} uses multiplications of operands quantized to different 
sizes, while the results are accumulated in 32-bit registers to 
avoid overflows and efficiently map to \ac{fpga} resources such as \acp{dsp}, as discussed in \Cref{sec:meth:conv}.

The quantization $Q(\cdot)$ of a  value $b$ on $\mathit{bw}$ bits
\begin{equation}
     a = Q(b) = \mathrm{clip}(\mathrm{round}(b \cdot 2^{\mathit{bw}-s}), a_\mathrm{min}, a_\mathrm{max}) \cdot 2^s  \quad s \in \mathbb{N}
     \label{eq:quant:0}
\end{equation}
\begin{align}
     a_\mathrm{min} = \mathit{act}_\mathrm{min}(s) = \Biggl\{
        \begin{array}{ll}
            0 & \text{if unsigned} \\
            -2^{\mathit{bw}-1-s} & \text{if signed}
        \end{array}
     \label{eq:quant:1}
\end{align}
\begin{align}
     a_\mathrm{max} = \mathit{act}_\mathrm{max}(s) = \Biggl\{
         \begin{array}{ll}
            2^{\mathit{bw}-s}-1 & \text{if unsigned} \\
            2^{\mathit{bw}-1-s} & \text{if signed}
        \end{array}
     \label{eq:quant:2}
\end{align}
uses $\mathit{s}$ as scaling factor, $a_\mathrm{min}$ is the 
lower clipping bound and $a_\mathrm{max}$ is the higher one. All 
\emph{zero points} are set to zero and omitted in the 
expressions above, while the \emph{scaling factors} are set to 
powers of two to map alignment operations between weights and 
activations into hardware-friendly bit shifts \cite{jin2022}.

The bias scaling factor $s_\mathrm{b}$ is calculated as the sum 
of the input scaling factor $s_\mathrm{x}$ and the weight scaling 
factor $s_\mathrm{w}$.

After the training, to avoid the hardware implementation
of floating point operations, the batch normalization layers 
are merged with the quantized convolution layers \cite{quant2018} 
and retrained to calibrate and tune the quantization parameters.
The final model is exported to the QONNX format 
\cite{Pappalardo2022,umuroglu2023}.

\subsection{Accelerator architecture}
\label{sec:meth:net}

The \emph{code generation} step in \cref{fig:flow} works on the 
optimized QONNX graph, i.e. after ReLU and batch normalization 
were merged with convolutional layers, and provides a C++ \emph{top 
function} that instantiates all the tasks (also known as dataflow processes) needed to implement 
network inference:
\begin{figure*}
\centering
\includegraphics[width=.8\textwidth]{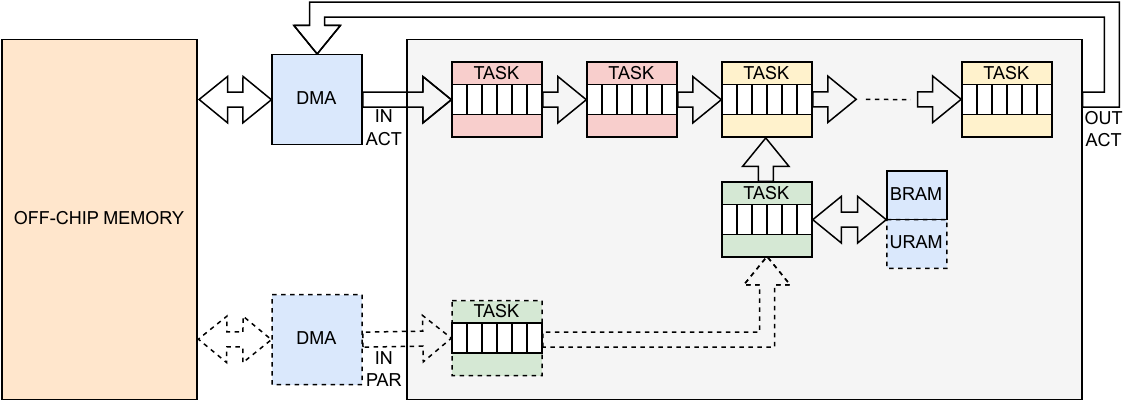}
\caption{\label{fig:arch} Accelerator architecture with 
\acf{dma} blocks for memory transfers (grey box) and
concurrent tasks [computation (yellow), buffer (red), and 
parameter (green)] communicating through data streams. Parameter 
loading from off-chip memory to URAMs (dashed) can be enabled on 
platforms supporting it.}
\end{figure*}
\begin{itemize}
    \item \emph{Computation task}: one for each convolution or 
    pooling node to implement the layer computations (see 
    \cref{sec:meth:conv}).

    \item \emph{Parameter task}: one for each convolution to 
    correctly provide the convolution parameters to the computation 
    task. One additional task is added to load parameters from 
    \emph{off-chip memory} if \ac{uram} storage is used (see 
    \cref{sec:meth:params}).

    \item \emph{Window buffer tasks}: multiple tasks for each 
    convolution or pooling node, to format the input data to 
    the computation tasks (see \cref{sec:meth:wb}).

\end{itemize}

All tasks are coded in a reusable, templated C++ library that can 
adapt to different activation and weight tensor dimensions, data 
types, and computational parallelism. Each task is composed of a 
main loop that performs multiple operations.


To increase the accelerator throughput, pipelining is enabled 
at two different levels:
\begin{itemize}
    \item Inter-task: concurrent layer execution is achieved 
    using the \emph{dataflow} pragma in the body of the top 
    function. There is one computation task and, possibly, 
    multiple window buffer tasks running for each of the network 
    layers. The latency of the slowest task determines the overall 
    accelerator throughput.

    \item Intra-task: concurrent operation execution is 
    used to reduce and balance task latencies. 
    \emph{Computation tasks} are the most computationally intensive. Thus 
    each top loop inside them is partially unrolled by a factor computed at compile 
    time and based on the complexity of the corresponding 
    computation. This effectively allocates a number of \acp{pe}, one for each unrolled iteration, for 
    each \emph{computation task}. Each \ac{pe} 
    performs one or more \acp{mac} operations per clock cycle. 
    See \cref{sec:meth:conv} and \cref{sec:meth:thopt} about how low-level \ac{dsp} packing is used to increase the number of \acp{mac} executed by a \ac{dsp} unit in a clock cycle. If the 
    \emph{computation task} belongs to a convolution, the related 
    \emph{parameter task} main loop is unrolled by the same 
    factor, to provide enough data to support the increased computations parallelism.
\end{itemize}
An \ac{ilp} model described in \cref{sec:meth:thopt} is used to globally optimize the unroll factors (number of \acp{pe}) that maximizes \ac{nn} throughput under \ac{dsp} resource constraints (\acp{dsp} are the most critical \ac{fpga} resource for the \ac{nn} architectures that we considered in this paper).

Network inputs and outputs are implemented as \emph{streams} of 
data. \ac{dma} blocks read/write input/output tensors to/from the off-chip memory. Streams also transfer data 
between tasks in order minimize the memory requirements for 
on-chip activations storage. (See \cref{sec:meth:wb})

The \emph{data-driven} execution approach is chosen to process 
the frames sequentially and as a continuous stream. This is 
achieved in Vitis HLS by using the \texttt{ap\_ctrl\_none} pragma in the top 
function that models the entire \ac{nn}. Each task is then operating as soon as input data are 
available.

Inference begins as soon as the \ac{dma} attached to the input 
port of the top-level interface is enabled to download input 
images. Each task is pipelined; the first stage reads 
the input stream, while the others process the data and write the 
output stream. As a further, tool-specific, implementation detail, intra-task pipelines are not flushable, which would consume more resources, but stalling  with auto-rewind disabled, to both save resources and avoid 
deadlocks. Auto-rewind would start a new loop execution while the 
pipeline is processing the last iterations of the old one, but with data-driven \texttt{ap\_ctrl\_none} dataflow it would cause deadlocks at runtime. Performance is largely unaffected because the \emph{intra-task 
pipeline} latency is very small, just a few cycles, compared to 
the task latency, which is proportional to the number of iterations of the 
\emph{intra-task pipeline}.


\subsection{Convolution computation task}
\label{sec:meth:conv}

\begin{figure*}
\centering
\includegraphics[width=\textwidth]{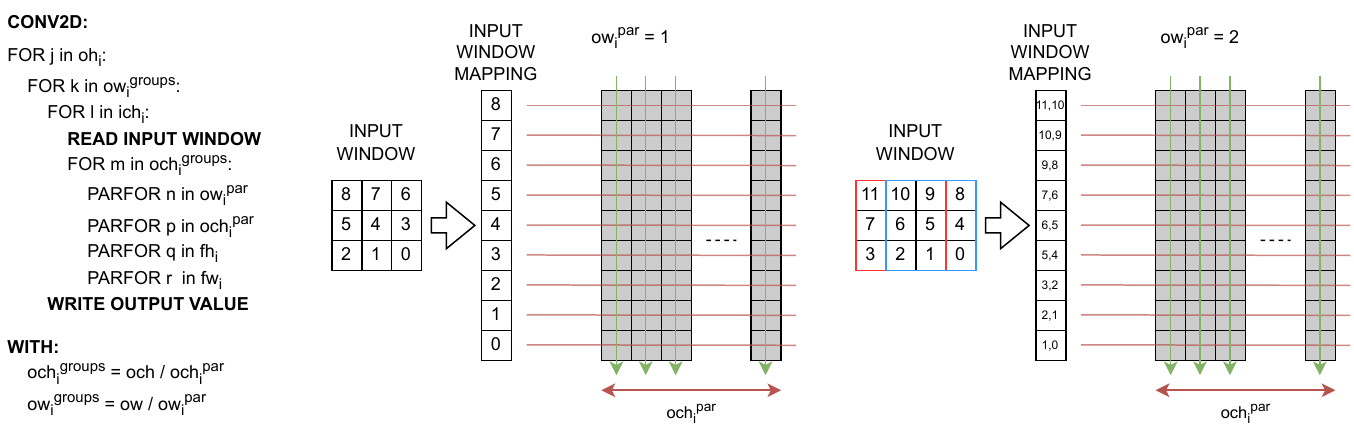}
\caption{Convolution architecture: The data flow is output 
stationary, and for $\mathit{och}^\mathit{par}_i$ output channels 
the contribution of the input window $\mathit{fh}_i, 
\mathit{fw}_i$ is evaluated every clock cycle. Data is written to 
the output after all input channels have been processed. The 
dataflow setup for $\mathit{ow}^\mathit{par}_i=1$ and 
$\mathit{ow}^\mathit{par}_i=2$ is shown in the two schematics. 
The input activations are loaded simultaneously, along orange 
lines, into each gray box, which is a \ac{pe}. \acp{pe} performs 
a \textit{MAC} operation and the partial results move through the 
pipeline from top to bottom along the green lines.}
\label{fig:conv_comp}
\end{figure*}

Each convolution \emph{computation task} receives a window of 
input activations from a \emph{window buffer task}. 
\cref{fig:conv_comp} shows the pseudo-code for the convolution 
computation and examples of how the computation pipeline receives 
input data and computes the partial results. The PARFOR 
pseudo-code is implemented as an unrolled loop in synthesizable 
C++.
\begin{table}
    \caption{Symbol definitions for layer $i$}
    \label{tab:dims}
    \centering
    \small
    \begin{tabular}{lcl}
         \toprule
         Symbol
            & \phantom{aa}
                & Description \\
         \midrule
         $\mathit{ich}_i$  && Input tensor channels  \\
         $\mathit{ih}_i$   && Input tensor height    \\
         $\mathit{iw}_i$   && Input tensor width     \\
         $\mathit{och}_i$  && Output tensor channels \\
         $\mathit{oh}_i$   && Output tensor height   \\
         $\mathit{ow}_i$   && Output tensor width    \\
         $\mathit{fh}_i$   && Filter tensor height \\
         $\mathit{fw}_i$   && Filter tensor width \\
         $\mathit{s}_i$    && Convolution stride \\
         \bottomrule
    \end{tabular}
\end{table}

Input tensors are mostly provided in depth-first-order to each 
convolution $i$, as discussed below. The innermost loops are completely 
unrolled over the filter dimensions ($\mathit{fh}_i, 
\mathit{fw}_i$) and part of the output channel ($\mathit{och}_i$) 
dimension. This unroll factor $\mathit{och}^\mathit{par}_i$, where ``par'' means that the execution will be fully data-parallel, defines the number of \acp{pe}, as discussed above. It is chosen 
by the algorithm described in \cref{sec:meth:thopt}. The 
$\mathit{och}^\mathit{par}_i$ unroll factor is limited by on-chip 
memory bandwidth and the number of arithmetic units that can be 
used simultaneously. Increasing the number of output channels 
computed in parallel per clock cycle requires the corresponding 
filter parameters to be provided in parallel, i.e., higher memory 
bandwidth and potentially more \ac{bram} resources.

Another optimization changes the order in which the windows are 
given to the data path, instead of channel first order, and 
unrolls along the output tensor width ($\mathit{ow}_i$) loop by a 
factor ($\mathit{ow}^\mathit{par}_i$). Unrolling along the tensor 
width allows us to reduce the computation time without requiring 
more memory bandwidth for the filter parameters, at the cost of 
more partitioning of the input activation window buffer, and hence of potentially more \ac{bram} resources.

This also allows the weights to be reused within an output 
stationary dataflow and can be exploited in future work where 
larger networks are considered and the off-chip memory is used to 
store network parameters. 

We now discuss how we exploit the \ac{dsp} packing method 
described in \cite{xilinxwp} to reduce the 
hardware overhead of computing quantized data, by performing 
multiple operations on a single \acp{dsp} block. Unlike 
$\mathit{och}^\mathit{par}_i$, which is resource dependent, 
$\mathit{ow}^\mathit{par}_i$ depends on the activation 
quantization bits.
Even though the number of operations packed in a \ac{dsp} depends 
on the number of bits, this work only used the configuration 
described in \cite{xilinxwp}, which presents a method for 
$\mathit{bw}_i=8$ for both 
parameters and activations.

\begin{figure}
    \centering
    \includegraphics[width=0.7\columnwidth]{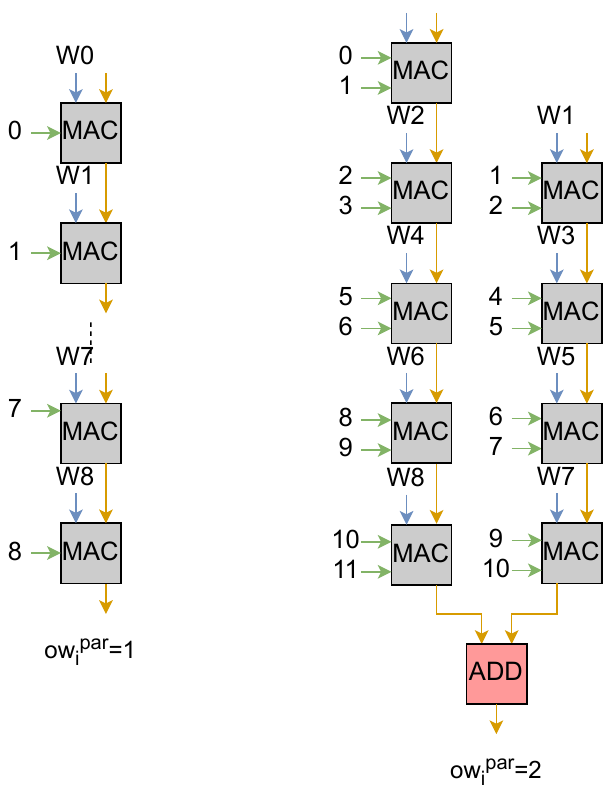}
    \caption{
    For $\mathit{ow^{par}}_i = 1 \implies$ pipeline stages $= 
    \text{\# of \acp{mac}} = \mathit{fh}_i \cdot \mathit{fw}_i$,
    because the maximum number of chainable operations is much higher.
    for $\mathit{ow^{par}}_i = 2$ (max for \SI{8}{bit} operands), 
    the pipeline is split in 2 (packing reduces max accumulations 
    per chain). An ADD combines the results.
    }
    \label{fig:pipe}
\end{figure}

\cref{fig:pipe} shows two examples of calculation pipelines, with 
different values of $\mathit{ow}^\mathit{par}_i$.

Each gray box is a \ac{pe} that receives:
\begin{itemize}
    \item \emph{Input activations}: $\mathit{ow}^\mathit{par}_i$ 
    inputs. These values change at each iteration of the 
    $\mathit{och}^\mathit{groups}_i$ loop. The input activations 
    are multiplied in parallel by the \ac{pe} input weight and 
    are provided by the corresponding \emph{window buffer tasks}.

    \item \emph{Input weight}: one input. This value is updated at 
    each clock cycle. The input weight is provided by the 
    corresponding \emph{parameter task}.

    \item \emph{Partial accumulation}: one input. This value 
    is updated at each clock cycle. The partial accumulation is 
    provided by the previous pipeline stage.
\end{itemize}

Each \ac{pe} receives an input weight every clock cycle, so 
sufficient \ac{ocm} bandwidth must be provided (see 
\cref{sec:meth:params}).

The two pipelines in \cref{fig:pipe} highlight how 
$\mathit{och}^\mathit{par}_i$ allocates multiple \acp{pe} per 
pipeline stage (horizontal unroll), and how 
$\mathit{ow}^\mathit{par}_i = 2$ modifies the mapping of the 
input activations to the different stages of the pipelines, thus 
increasing the number of computations for each \ac{pe}. The 
partial accumulation entering each \ac{pe} comes from the 
previous pipeline stage. The only exception is the first stage, 
which receives as value to accumulate the \emph{bias} of the 
convolution.


Each \ac{mac} calculation, for the case 
$\mathit{ow}^\mathit{par}_i = 1$, is done using a \ac{dsp} from 
the \emph{Xilinx} architecture. If $\mathit{ow}^\mathit{par}_i = 
2$ the two \acp{mac} have reduced resource usage thanks to the 
technique described in \cite{xilinxwp}.

As shown by the pipeline in \cref{fig:pipe}
with $\mathit{ow}^\mathit{par}_i = 2$, the operation packing is 
done by multiplying 2 activations ($A$, $D$) by 1 parameter ($B$) 
and accumulating to a partial result ($P_{i-1}$). The output 
($P_i$) is passed to the next pipeline stage.

The multiplier of the \acp{dsp} receives one \SI{27}{bit} and one 
\SI{18}{bit} input. The former packs the two activations:
\begin{center}
\includegraphics[width=.85\columnwidth]{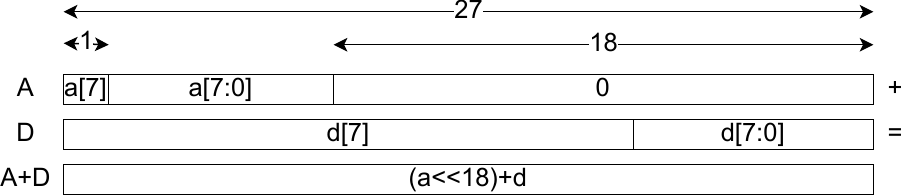}
\end{center}
%
while the latter contains the  sign-extended parameter:
\begin{center}
\includegraphics[width=.55\columnwidth]{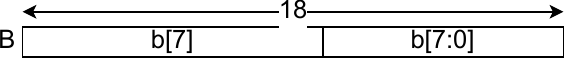}
\end{center}
The two operands are multiplied into a \SI{36}{bit} word ($M$):
\begin{center}
\includegraphics[width=\columnwidth]{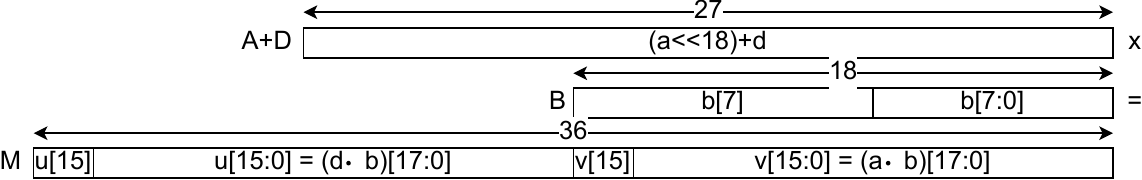}
\end{center}
The \ac{dsp} also adds the partial product to the accumulation coming 
from the previous pipeline stage, $P_{i-1}$:
\begin{center}
\includegraphics[width=\columnwidth]{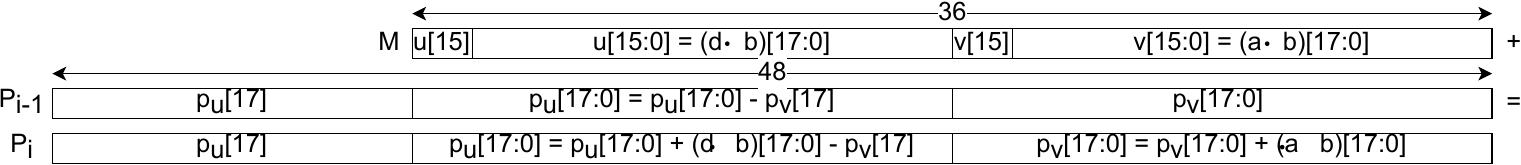}
\end{center}
At the end of the chain, a restore stage ensures that the $p_v$ 
sign does not create errors in the final result:
\begin{center}
\includegraphics[width=\columnwidth]{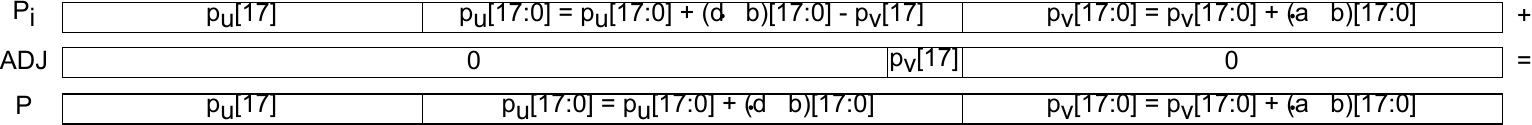}
\end{center}

Note that in our specific use case, where activations and parameters are quantized on \SI{8}{bits}, we can at most chain {\bf 7} packed \acp{dsp}, because of 
the limited padding between the two words, namely \SI{2}{bits}, and the restore mechanism which corrects \SI{1}{bit} overflow. 
However, for convolution filters with $\mathit{fh}_i = 3$ and $\mathit{fw}_i = 
3$, the \ac{dsp} chain should have a length of 9.
Hence we split the chain into \emph{2} subparts that respect the maximum length condition. 
The partial accumulations coming from the different chains are then added together in an additional stage,
and the result coming from the \acp{dsp} pipeline is finally added to the accumulation register.


Registers keeping the partial results of the \ac{mac} between multiple iterations are sized in order to avoid overflows in the computations. 
Considering models with $\mathit{bw}_i$ bit quantization, the 
accumulated values from the product between an activation and a parameter, have a width equal to 
$2 \cdot \mathit{bw}_i$, using same bit-width for activation and 
parameter quantization.
For each convolution, the number of accumulations ($N^{acc}_{i}$) 
performed to compute each output value is
\begin{equation}
    N^\mathit{acc}_i = \mathit{och}_i \cdot \mathit{ich}_i \cdot \mathit{fh}_i \cdot \mathit{fw}_i.
\end{equation}
Since the addend has $2 \mathit{bw}_i$ bits, the final 
accumulation register must have a width equal to
\begin{equation}
    \mathit{bw}^\mathit{acc}_i = \lceil \log_2(N^\mathit{acc}_i) \rceil + 2 \mathit{bw}_i.
\end{equation}
Considering the worst case for \emph{Resnet8} and \emph{Resnet20} 
with \SI{8}{bit} quantization, the required bitwidth is
\begin{equation}
    N^\mathit{acc}_i = 32 \cdot 32 \cdot 3 \cdot 3 = 9216
\end{equation}
\begin{equation}
    \mathit{bw}^\mathit{acc}_i = \lceil \log_2 9216 \rceil + 2 \cdot 8 = 14 + 16 = 30.
\end{equation}
The accumulation register size is chosen to be \SI{32}{bit} because it ensures no overflow, and using standard C++ types improves C simulation speed.

\subsection{Parameter task}
\label{sec:meth:params}

Each convolution layer of the QONNX network graph has a 
\emph{parameter task} in the top function, feeding the 
computation pipeline with data from on-chip memory. Depending on 
the target \ac{fpga}, parameters may be stored in:
\begin{itemize}
    \item
    \emph{\Acp{bram}}: they can store up to \SI{4}{KB} each and can be 
    initialized by the bitstream. The parameters for each convolution 
    are stored in separate arrays, one for each weight and bias 
    of the convolutions, because each is accessed by a specific 
    \emph{parameter task}.

    \item \emph{\acfp{uram}}: they can store \SI{32}{KB} of data each 
    (allowing higher performance) but require dedicated hardware 
    for initialization (a \ac{dma}-driven input stream). 
    The parameters for each convolution are packed into a single array stored in 
    off-chip DRAM (also accessible by the host) and transferred by
    \ac{dma} once at power-up. A concurrent task in 
    the dataflow architecture splits and distributes the input 
    parameter stream to the tasks that handles the parameters of 
    each convolution. Each \emph{Parameter task} provides the filter 
    data to the computation pipeline and caches it in \acp{uram} 
    at the first iteration for reuse (hence the next \acp{uram} 
    accesses are read-only).

\end{itemize}

The Ultra96 board lacks \ac{uram}, so \ac{bram} is used. The KRIA 
KV260 board uses the \ac{uram}.


As discussed in \cref{sec:meth:conv}, the main loop of each 
convolution's \emph{computation task} consumes $\mathit{cw}_i = 
\mathit{och}^\mathit{par}_i \cdot \mathit{fh}_i \cdot 
\mathit{fw}_i$ filter data per clock cycle. The 
$\mathit{ow}^\mathit{par}_i$ unroll factor does not contribute 
because each parameter is used for multiplication with 
$\mathit{ow}^\mathit{par}_i$ activations. To avoid stalling the 
computation pipeline, the \emph{parameter task} must write 
$\mathit{cw}_i$ weights every clock cycle and read the same 
amount of data from the \acp{bram} or \acp{uram}. Arrays are then 
reshaped by a factor equal to $\mathit{cw}_i$, using the 
\texttt{array\_reshape} pragma, to achieve the required memory 
bandwidth.

\subsection{Throughput optimization}
\label{sec:meth:thopt}

To avoid stalling, all streams are sized appropriately by our configuration 
\emph{Python} script based on their type, as follows.

Streams created by \emph{parameter tasks} supply 
\emph{computation tasks} with a token size equal to the 
computational parallelism of the consuming convolution, 
$\mathit{och}^\mathit{par}_i$, every clock cycle. Since the 
producer and consumer write and read one token per clock cycle, 
the stream size is 2.

The sizes of the streams produced by \emph{window buffer tasks} 
are discussed in \cref{sec:meth:wb}.

The output stream from \emph{computation tasks} must consider 
$\mathit{och}^\mathit{par}_i$ and $\mathit{ow}^\mathit{par}_i$. 
The pseudocode in \cref{fig:conv_comp} shows that 
\emph{computation tasks} write a burst of $\mathit{och}_i \cdot 
\mathit{ow}^\mathit{par}_i$ output activations, grouped into 
tokens of size $\mathit{och}^\mathit{par}_i$ to not stall the 
pipeline. When packing is applied, the output stream is split 
into $\mathit{ow}^\mathit{par}_i$ parallel channels
to ensure enough bandwidth.
Each channel is implemented by a \ac{fifo} of size $\mathit{och}^\mathit{groups}_i = 
\mathit{och}_i/\mathit{och}^\mathit{par}_i$ to store the burst 
transactions completely.


As mentioned above, using the \emph{dataflow} paradigm and assuming optimal stream 
sizing to avoid stalling, accelerator throughput is limited by 
the slowest concurrent process. Therefore, the throughput 
$\mathit{Th}$ of each layer unit must be balanced for optimal 
performance. The latency of each module depends on the number of 
computations for each input frame $c$ and the computational 
parallelism $\mathit{cp}$ required for each block $i$. The number 
of computations for a convolutional layer is
\begin{equation}
    c_i = \mathit{oh}_i \cdot \mathit{ow}_i \cdot \mathit{och}_i \cdot \mathit{ich}_i \cdot \mathit{fh}_i \cdot \mathit{fw}_i.
    \label{eq:ilp:0}
\end{equation}
Since the parameter $c_i$ is fixed and depends on the chosen 
network architecture, the throughput per layer is set by the 
number of compute units allocated to each \emph{computation task} 
implementing a layer.
As shown in the pseudocode in \cref{fig:conv_comp}, computation 
parallelism $\mathit{cp}_i$ is
\begin{align}
    \mathit{cp}_i &= k_i \cdot \mathit{och}^\mathit{par}_i \cdot \mathit{ow}^\mathit{par}_i, \\
    \text{with} \quad k_i &= \mathit{fh}_i \cdot \mathit{fw}_i, \mathit{och}^\mathit{par}_i, \mathit{ow}^\mathit{par}_i \in \mathbb{N}.
\end{align}
Since the filter size $k_i$ is defined by the model and 
$\mathit{ow}^\mathit{par}_i = 2$, because for this work we 
consider all the quantization bit-widths equal to \SI{8}{bit}, 
the variable to optimize is $\mathit{och}^\mathit{par}_i$, i.e. 
$\mathit{cp}_i$ is an integer multiple of the filter size.

The throughput of each task, $\mathit{Th}_i$ \ac{fps}, depends on 
the variable $\mathit{och}^\mathit{par}_i$
\begin{equation}
    \mathit{Th}_i = \mathit{Th}\left(\mathit{och}^\mathit{par}_i\right) = \frac{\mathit{cp}_i}{c_i} = \frac{k_i \cdot \mathit{och}^\mathit{par}_i \cdot \mathit{ow}^\mathit{par}_i}{c_i}.
    \label{eq:ilp:2}
\end{equation}

Considering a network with $N$ convolutional layers, \cref{alg:0} 
\begin{algorithm}[tb]
\caption{ILP Model}
\label{alg:0}
\small
\textbf{Inputs:} $i_\mathrm{max}$, $N_\mathrm{PAR}$ \\ 
\textbf{Variable:} $\mathit{och}^\mathit{par}_{i_\mathrm{max}}$
\begin{alignat}{2}
  \text{maximize}   & \quad & \mathit{Th}\left(\mathit{och}^\mathit{par}_{i_{max}}\right) \label{eq:ilp:5}\\
  \text{subject to} &       & \mathit{cp}_\mathrm{tot} = \sum_{i = 1}^N{\mathit{cp}_{i_\mathrm{max}} \cdot r_i} \leq N_\mathrm{PAR} \label{eq:ilp:6}
\end{alignat}
\end{algorithm}
shows \iac{ilp} formulation of throughput optimization. If 
$i_\mathrm{max} \in [1,N]$ is the index of the layer with the 
highest $c_i$, then the goal is to balance the throughput of all 
layers
\begin{equation}
    \forall {i \in [1,N]} \quad {\mathit{Th}\left(\mathit{och}^\mathit{par}_{i_\mathrm{max}}\right) = \mathit{Th}\left(x_i\right)} \implies \mathit{cp}_i = \mathit{cp}_{i_\mathrm{max}} r_i
    \label{eq:ilp:3}
\end{equation}
with $r_i = c_i / c_{i_\mathrm{max}}$. Then the number of 
resources needed for each layer can be calculated, given the 
resources allocated for layer $i_\mathrm{max}$. The total number 
of parallel computations allocated is
\begin{equation}
    \mathit{cp}_\mathrm{tot} = \sum_{i = 1}^N{\mathit{cp}_i} = \sum_{i = 1}^N{\mathit{cp}_{i_\mathrm{max}} r_i}.
    \label{eq:ilp:4}
\end{equation}

From \cref{eq:ilp:6}, $N_\mathrm{PAR}$ limits the maximum number 
of computations that can be done in parallel and depends on the 
platform. The \acp{fpga} on the Ultra96 and KRIA KV260 boards that we 
are considering have \num{360} and \num{1248} \acp{dsp}, 
respectively. During hardware generation, $N_\mathrm{PAR}$ is set 
to the number of \acp{dsp} on the target board.


The \acs{ilp} can then maximize the throughput of the network by 
optimizing the parameters for the $i_\mathrm{max}$ layer 
\cref{eq:ilp:5} and automatically configuring the template 
parameters of the tasks.

\subsection{Window generation}
\label{sec:meth:wb}

\begin{figure}
    \centering
    \includegraphics[width=.7\columnwidth]{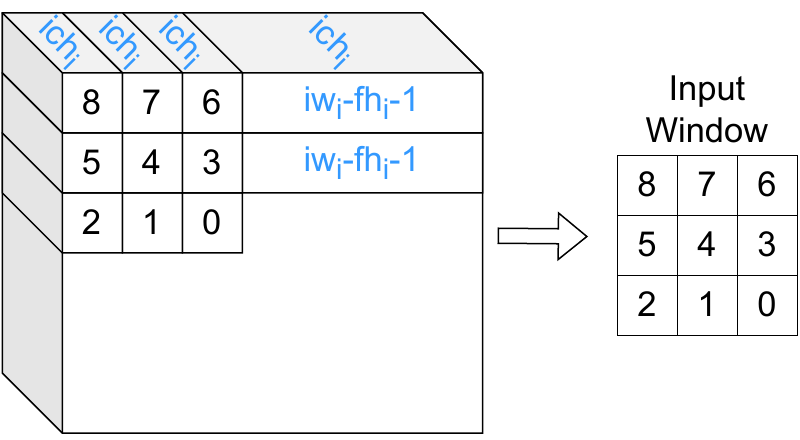}
    \caption{Input window mapped on the input tensor, $\mathit{ow^{par}}_i = 1$.
    \cref{fig:conv_comp} shows how the window elements map to the computation pipeline.}
    \label{fig:window_buffer:iact0}
\end{figure}

Given a convolution input tensor, we only need to 
store on-chip enough data to provide the input window to the 
\emph{intra-task} pipeline of each computational task. For 
example, \cref{fig:window_buffer:iact0}
shows an input tensor and the input window mapping for a 
convolution with $\mathit{fh}_i = 3, \mathit{fw}_i = 3$. 
It is important to highlight that the activations are produced using \emph{depth-first} order by the convolution that creates the input tensor (\cref{fig:conv_comp}), while the input window is distributed over one channel and multiple lines.
It is thus necessary to store all the lines needed to generate an input window, so each window buffer 
(also called line buffer in the literature) should be sized to 
accommodate the required activations. The portion of the input 
tensor ($\mathit{B_{i}}$) that the buffer must retain to create 
an input window is highlighted in \cref{fig:window_buffer:iact0}
\begin{equation}
    B_i = \left[\left(\mathit{fh}_i-1\right) \cdot \mathit{iw}_i + \mathit{fw}_i -1\right] \cdot \mathit{ich}_i.
    \label{eq:lb:0}
\end{equation}

This size is constant over time because each time that the buffer reads an activation and generates a window, it discards the older value (\cite{lb2019}).

The \emph{window buffer tasks} retrieve from the input buffer the 
$\mathit{B_i}$ activations required for the windows. At the maximum unroll 
factor, $\mathit{och}^\mathit{par}_i = och_i$, each intra-task pipeline of the \emph{computation task} processes one 
input window per clock cycle.

\begin{figure}[H]
    \centering
    \includegraphics[width=\columnwidth]{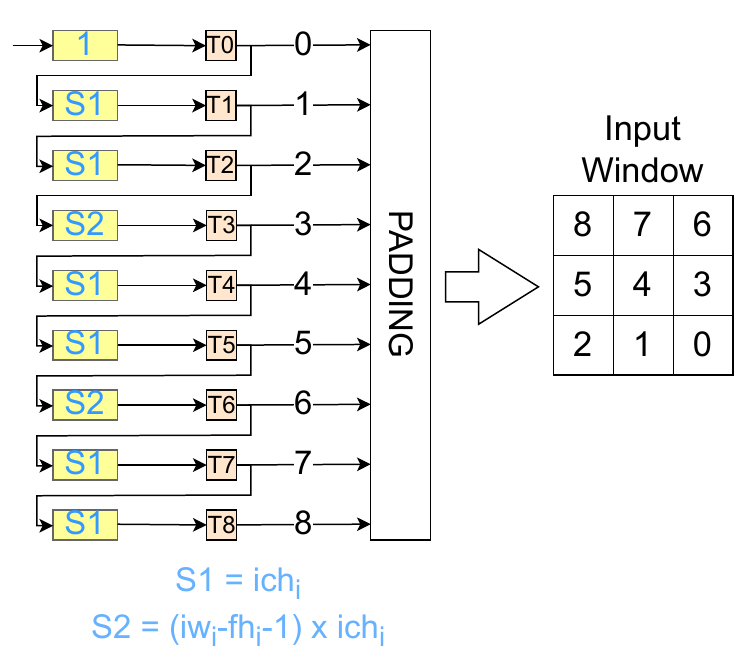}
    \caption{
    Buffer partitioning, $\mathit{ow}^\mathit{par}_i = 1$. Yellow 
    boxes are the \acp{fifo} with their sizes. Orange boxes are 
    tasks that read and write the \acp{fifo}. Padding is applied 
    before generating the box for the convolution.
    }
    \label{fig:window_buffer:iact1}
\end{figure}

The data read by the \emph{window buffer tasks} from the input 
activation buffer is $\mathit{fh}_i \cdot \mathit{fw}_i$, i.e. a 
convolution window. The data needed for the input window is not 
contiguous and cannot be read by directly addressing the buffer, 
because it is stored sequentially in a \ac{fifo} with only one 
read port available. To provide the necessary bandwidth, the 
\ac{fifo} must be partitioned into $\mathit{fh}_i \cdot 
\mathit{fw}_i$ parts, connected sequentially as shown in 
\cref{fig:window_buffer:iact1}. Optimizing the window buffer to 
reduce the required partitioning in cases that allow a lower 
window generation rate is left for future work.

The size of each \ac{fifo}, $S_1$, $S_2$, is the distance, in 
number of activations, between successive values of the same 
input window, considering that the tensor is processed in
\emph{depth-first} order. 
$\mathit{S}_1$ represents the distance between two activations within the same row of the input window, and it is equal to the number of channels $\mathit{ich}_i$ in the tensor. 
In contrast, $\mathit{S}_2$ covers the gap between two activations in different rows of the input window, and it is directly proportional to one row ($\mathit{ich}_i \cdot \mathit{iw}_i$) in the input tensor, minus the filter width $\mathit{fw}_i$.
Each \emph{task} $T_i$ reads from a 
\ac{fifo} the data for an input window position, $i$, and 
provides the input for the next \ac{fifo} slice, $i+1$.

The \emph{padding} task, if enabled, reads at each cycle the data 
from \emph{task} $T_i$ for positions $i$ that do not require padding, and replaces with \num{0} the positions of the 
input window that must be padded. Thanks to the 
concurrent execution and padding-aware control of the \emph{window 
buffer tasks} and \emph{padding task}, the first buffer slices 
can be initialized with the tensor values of the next frame, 
while the last ones generate the final windows of the previous 
frame, avoiding the latency overhead caused by initializing the 
\acp{fifo}.

\begin{figure}
    \centering
    \includegraphics[width=\columnwidth]{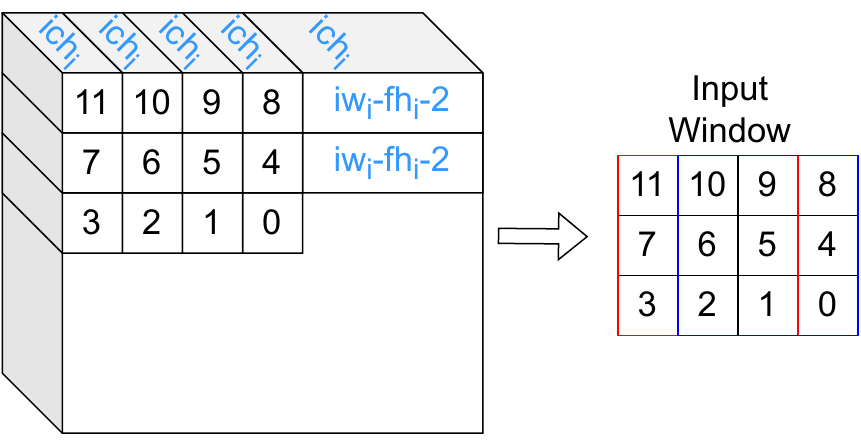}
    \caption{
    Input window mapped on the input tensor, 
    $\mathit{ow}^\mathit{par}_i = 2$, retaining two computation 
    windows (red and blue). \cref{fig:conv_comp} shows how the 
    window elements map to the computation pipeline.
    }
    \label{fig:window_buffer:iact2}
\end{figure}

The structure of the \emph{window buffer} depends 
on the unroll factor $\mathit{ow}^\mathit{par}_i$. 
\cref{fig:window_buffer:iact2} shows the input window mapped to 
the input tensor with $\mathit{ow}^\mathit{par}_i = 2$, for which 
all considerations made before about $\mathit{ow}^\mathit{par}_i = 
1$ apply.

\begin{figure}
    \centering
    \includegraphics[width=\columnwidth]{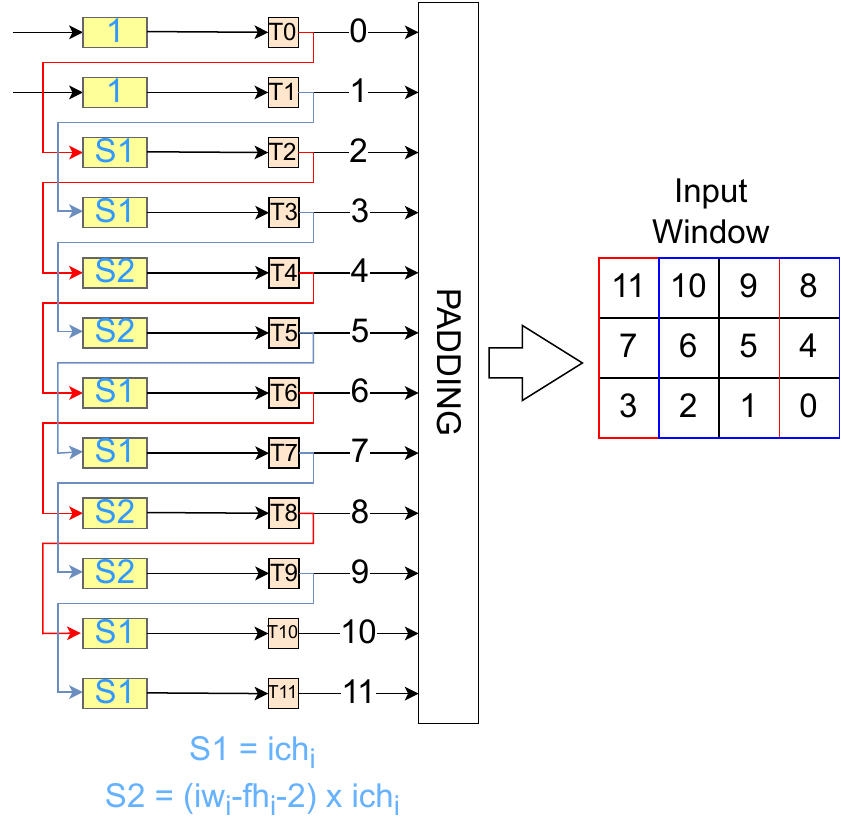}
    \caption{Buffer partitioning, $\mathit{ow}^\mathit{par}_i = 
    2$. The data in output from each \emph{task Ti} is connected 
    as input to the \ac{fifo} slice at the position 
    $\mathit{i+2}$ because of activation reuse.}
    \label{fig:window_buffer:iact3}
\end{figure}

The input buffer size is
\begin{equation}
    B_i = \left[\left(\mathit{fh}_i-1\right) \cdot \mathit{iw}_i + \mathit{fw}_i \right] \cdot \mathit{ich}_i
    \label{eq:lb:1}
\end{equation}
so the overhead with respect to \cref{eq:lb:1} is minimal. The 
buffer must be partitioned to ensure the required window 
production rate. With $\mathit{ow}^\mathit{par}_i = 2$, the 
elements of the input window are $(\mathit{fw}_i + 
\mathit{ow}^\mathit{par}_i - 1) \cdot \mathit{fh}_i$. 
\cref{fig:window_buffer:iact3} shows how the buffer is 
partitioned according to the required bandwidth.

The main difference between \cref{fig:window_buffer:iact1} and 
\cref{fig:window_buffer:iact3} is how the activations flow in the 
different \ac{fifo} slices.

Given an input filter of size $\mathit{fh}_i \cdot 
\mathit{fw}_i$, an input activation is multiplied, at different 
times, by a value in each position of the filter window.

If $\mathit{ow}^\mathit{par}_i = 1$, there is a one-to-one 
correspondence between the positions of the input window and 
those of the filter window. This means that the activation must 
pass through all \ac{fifo} slices, because each of them 
represents a position $i$ of the input/filter windows.

If $\mathit{ow}^\mathit{par}_i = 2$, the input window keeps two 
windows that are multiplied by the parameter window, i.e. part 
of the activations are evaluated in two adjacent positions for 
each input window ($i, i+1$). Thus, the output of $T_i$ must be 
connected to the input of the \ac{fifo} slice $i+2$ to ensure 
correct data flow.

\subsection{Graph Optimization}
\label{sec:meth:graph}

The main contribution of this paper is to provide a structured 
methodology to efficiently implement a residual block in a 
dataflow accelerator with concurrent processes.

\begin{figure}
    \centering
    \captionsetup[subfigure]{justification=centering}
    \begin{subfigure}{0.4\columnwidth}
    \centering
        \includegraphics[width=0.6\columnwidth]{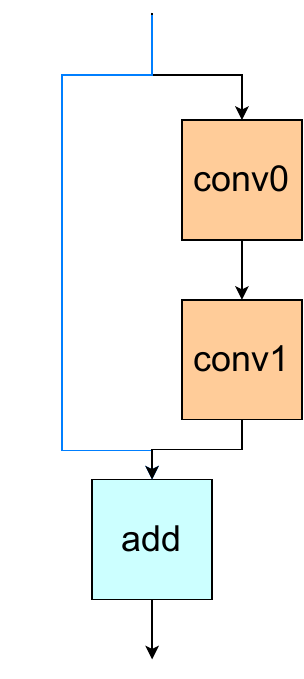}
        \caption{No downsample}
    \end{subfigure}
    \begin{subfigure}{0.4\columnwidth}
        \centering
        \includegraphics[width=0.6\columnwidth]{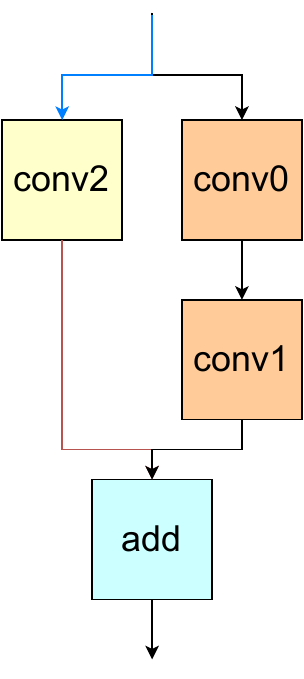}
        \caption{Downsample}
    \end{subfigure}
    \caption{\label{fig:meth:residual_block} \textit{Resnet20} and \textit{Resnet8} residual blocks}
\end{figure}

The same considerations from \cref{sec:meth:wb} can be extended 
to network graphs with multiple nodes processing the same input 
tensor, i.e. residual blocks, as shown in 
\cref{fig:meth:residual_block}, to provide a more general 
solution.
Considering a tensor processed by multiple convolutions, multiple 
branches start from the convolution that generates the tensor. 
In residual networks the branches are then merged by an 
\emph{add} layer.

\cref{fig:meth:residual_block} shows \textit{Resnet20} and 
\textit{Resnet8} residual block topologies with \emph{2 branches} 
per input tensor and \emph{0 or 1 convolutions} on skip 
connection (the branch crossing fewer convolutions).

The \emph{add} layer adds the values from the \emph{2 branches}. Because of the dataflow architecture, the operation 
starts as soon as both input streams have data. However, the 
time required to fill each stream is different. The skip 
connection stream that reaches the \emph{add} node is filled in 
parallel with the \emph{conv0} input stream in the case without 
downsampling, or after $\mathit{ich}_i$ cycles in the case with 
downsampling. The input stream from the long branch is filled as 
soon as \emph{conv1} provides its first output activation. As 
shown by \cref{fig:window_buffer:iact0}, \emph{conv1} starts 
processing data as soon as its input buffer is full. The amount 
of data buffered for skip connections, $B_\mathrm{sc}$, is equal 
to the amount to be processed by \emph{conv0} and is sufficient 
to start \emph{conv1} operations. To calculate this value we use 
the \emph{receptive field} \cite{araujo2019computing}, which is 
the portion of feature maps related to successive layers that 
contribute to produce an activation.

\begin{figure}
    \centering
    \includegraphics[width=0.9\columnwidth]{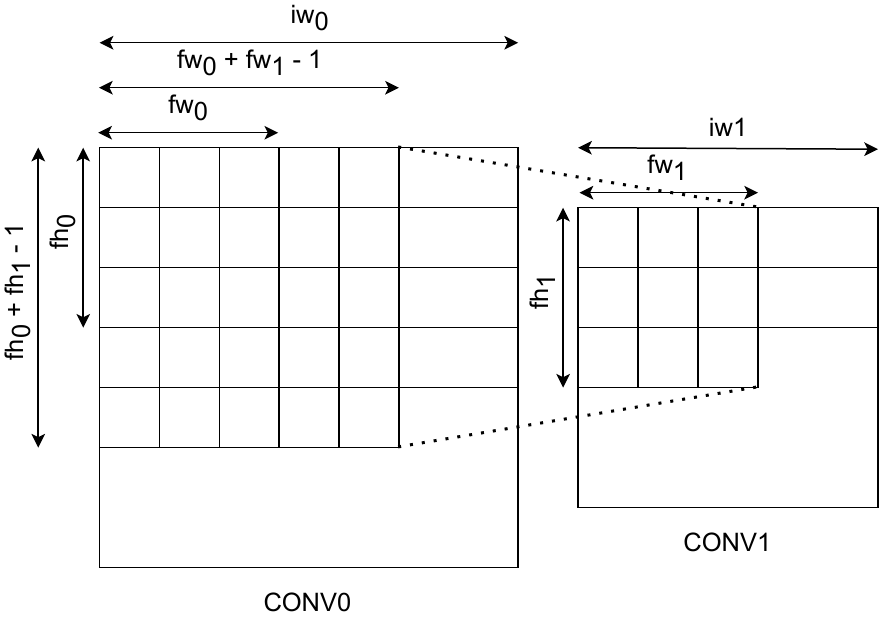}
    \caption{
    Receptive field of the \emph{conv1} window. For clarity, the 
    $\mathit{ich}$ dimension is omitted and \emph{conv1} stride 
    is assumed to be 1 ($s_1 = 1$).
    }
    \label{fig:meth:receptive_field}
\end{figure}

\cref{fig:meth:receptive_field} shows the \emph{receptive field} 
of the \emph{conv1} window with respect to the \emph{conv0} that 
generates it.
$B_\mathrm{sc}$ is the buffering of all receptive fields 
projected from the activation in the \emph{conv1} line buffer as 
soon as it starts computing. From \cite{araujo2019computing}, as 
shown in \cref{fig:meth:receptive_field}, the data to store for 
each receptive field $B_\mathrm{r}$ is
\begin{align}
    \mathit{rh}_0 &= \mathit{fh}_1 + \mathit{fh}_0 - 1 \\
    \mathit{rw}_0 &= \mathit{fw}_1 + \mathit{fw}_0 - 1 \\
    B_r &= \mathit{rh}_0 \cdot \mathit{rw}_0.
    \label{eq:rf:1}
\end{align}
Sliding the receptive field window over $\mathit{ich}_i, 
\mathit{iw}_i$, the obtained buffering $B_\mathrm{sc}$ is
\begin{gather}
    B_\mathrm{sc} = \left[\mathit{iw}_0 \left(\mathit{rh}_0 - 1\right) + \mathit{rw}_0\right] \mathit{ich}_0.
    \label{eq:rf:2}
\end{gather}


In the dataflow architecture used in the final implementation, 
the ``bypass'' branch must store its input activation data from 
the previous stage until the first output activation is generated 
by the convolution and the merged output can be generated. In 
previous dataflow implementations of \acp{cnn}, this buffering 
consumed large amounts of memory (\cite{weng2021hardwareefficient}). 

To efficiently support 
\emph{residual} blocks, the multiple endpoints of the input 
tensor and the increased buffering caused by the different number 
of convolutions (and thus different computation delays) per 
branch must be handled differently.

The combination of the following optimization steps for the 
dataflow architecture, \emph{proposed for the first time in this 
paper}, can avoid it, e.g., in \acp{cnn} such as Resnet8 and 
Resnet20.

The following two transformations (see 
\cref{fig:flow_merge_conv}) show how to solve the problem of 
multiple endpoints, reducing the length of the skip connection 
and the required buffering:
\begin{figure}
\captionsetup[subfigure]{justification=centering}
\centering
\phantom{}\hfill
\begin{subfigure}{0.4\columnwidth}
\centering
    \includegraphics[width=1\columnwidth]{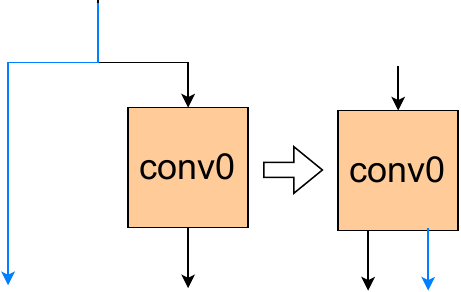}
    \caption{No downsample}
    \label{fig:flow_merge_conv_nodown}
\end{subfigure}
\hfill\hfill\hfill
\begin{subfigure}{0.46\columnwidth}
    \centering
    \includegraphics[width=1\columnwidth]{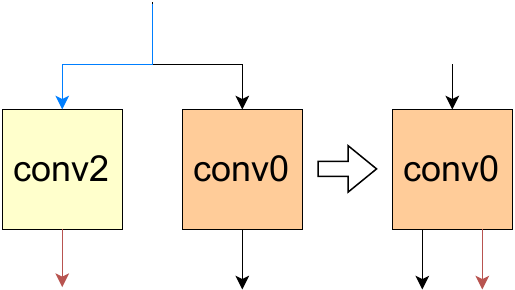}
    \caption{Downsample}
    \label{fig:flow_merge_conv_down}
\end{subfigure}
\hfill\phantom{}
\caption{
Multiple endpoint graph optimizations: 
(\subref{fig:flow_merge_conv_nodown}) input forwarding without 
downsampling, (\subref{fig:flow_merge_conv_down}) layer merging 
when there is a downsample pointwise convolution.
}
\label{fig:flow_merge_conv}
\end{figure}
\begin{itemize}
    \item
    Loop merge: if the residual block has a downsample layer, 
    i.e., a pointwise convolution in the short branch of the skip 
    connection, the algorithm merges the two convolution loops. 
    Both computations are performed by the same task, which 
    provides the tensor produced by the merged layer as an 
    additional output.

    \item
    Temporal reuse: to avoid buffering the same tensor twice, if 
    the residual block does not have a downsample layer, the 
    graph optimization uses the window buffer as input to the 
    convolution. The values are forwarded, using a second output 
    stream, to the next layer once they have been completely 
    used.

\end{itemize}

\begin{figure}
\centering
    \includegraphics[width=0.4\columnwidth]{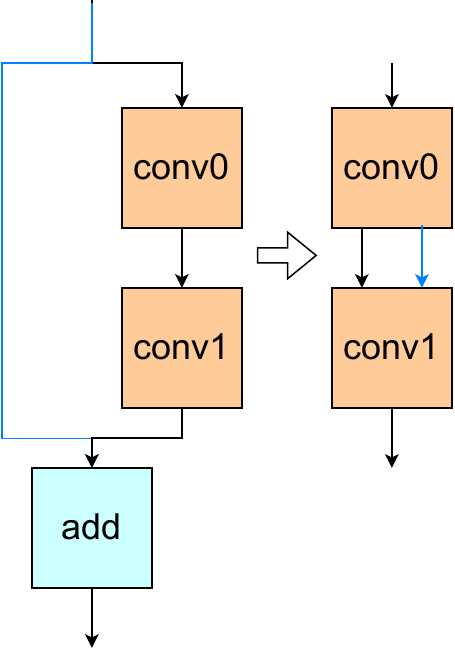}
    \caption{
    The addition is optimized as initialization of the 
    convolution accumulator register.
    }
    \label{fig:flow:2}
\end{figure}

Thanks to these transformations, the two input streams of the 
\emph{add} merge layer are produced simultaneously, and computation tasks never stall. \emph{Conv0} 
writes the skip connection stream as soon as the convolution 
computation starts and at the same rate as the convolution output 
tensor. The last transformation, shown in \cref{fig:flow:2}, 
removes the sum of the value coming from the short branch by 
connecting it as an additional contribution to the second 
convolution of the long branch. The value from the skip branch is 
used to initialize the accumulator register, and the addition 
operation is removed from the network graph.

The producer and consumer of the two branch streams are now the 
same (\emph{conv0} and \emph{conv1}), and they produce/consume at the same rate. 
The required buffering of the skip connection ($B_\mathrm{sc}$) 
is now equal to the \emph{conv1} window buffer size
\begin{equation}
    B_\mathrm{sc} = B_1 = \left[\left(\mathit{fh}_1-1\right) \cdot \mathit{iw}_1 + \mathit{fw}_1 -1\right] \cdot \mathit{ich}_1.
    \label{eq:lb:3}
\end{equation}

The dimensions of the first residual block without downsample of 
\emph{Resnet20} are: $\mathit{iw}_0 = \mathit{iw}_1 = 32$, 
$\mathit{ich}_0 = \mathit{ich}_1 = 16$, $\mathit{fh}_0 = 
\mathit{fh}_1 = 3$, $\mathit{fw}_0 = \mathit{fw}_1 = 3$.

The dimensions of the first residual block with downsample of 
\emph{Resnet20} are: $\mathit{iw}_0 = 32$,  $\mathit{iw}_1 = 16$, 
$\mathit{ich}_0 = 16$, $\mathit{ich}_1 = 32$, $\mathit{fh}_0 = 
\mathit{fh}_1 = 3$, $\mathit{fw}_0 = \mathit{fw}_1 = 3$.

The skip connection buffering, $B_\mathrm{sc}$, is then reduced 
to $R_\mathrm{sc}$, in both cases
\begin{gather}
    R_\mathrm{sc} = \frac{\left[\left(\mathit{fh}_1-1\right) \mathit{iw}_1 + \mathit{fw}_1 -1\right] \mathit{ich}_1}{\left[\left(\mathit{rh}_0 - 1\right) \mathit{iw}_0 + \mathit{rw}_0\right] \mathit{ich}_0} = 0.5.
\end{gather}

The same calculated gain holds for all residual blocks in 
\emph{Resnet20} because the product $\mathit{iw}_i \cdot 
\mathit{ich}_i$ remains constant. The same considerations apply 
to \emph{Resnet8}, since the structure of its residual blocks is 
identical to those already analyzed.

\begin{figure}
\captionsetup[subfigure]{justification=centering}
\centering
\phantom{}\hfill
\begin{subfigure}{0.4\columnwidth}
\centering
    \includegraphics[width=1\columnwidth]{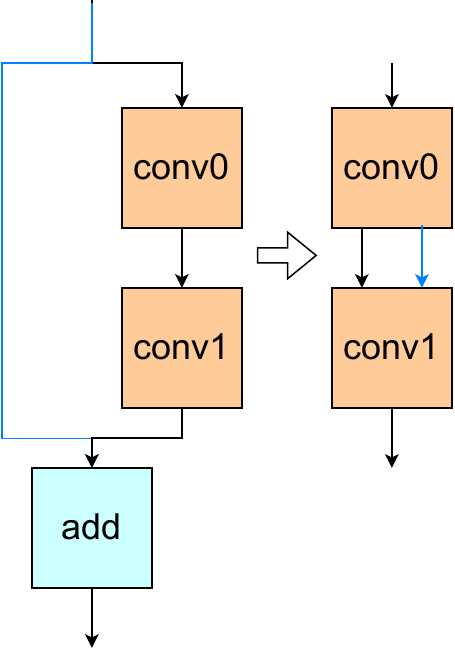}
    \caption{No downsample}
\end{subfigure}
\hfill\hfill\hfill
\begin{subfigure}{0.45\columnwidth}
    \centering
    \includegraphics[width=1\columnwidth]{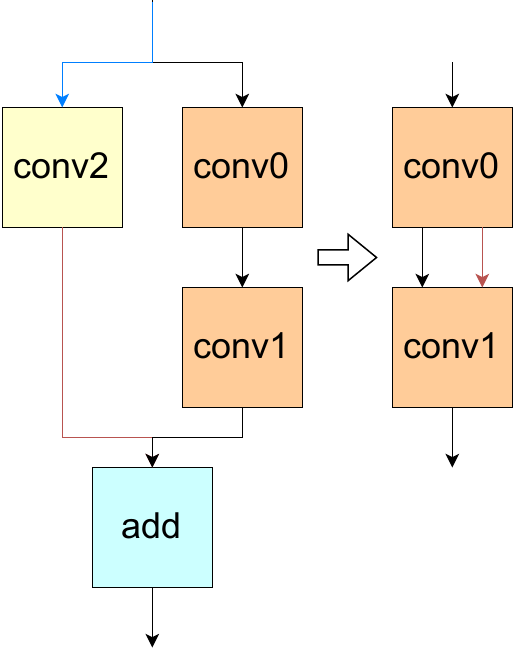}
    \caption{Downsample}
\end{subfigure}
\hfill\phantom{}
\caption{
Graph optimization for the residual blocks of Resnet8 and 
Resnet20 networks. The skip connection goes through the first 
convolution layer, conv0, into the second convolution layer, 
conv1, reducing buffering requirements with and without 
downsampling.
}
\label{fig:flow:3}
\end{figure}

From a network graph of a residual block with and without 
downsampling, \cref{fig:flow:3} shows the initial and final 
representations after applying the previously described 
optimizations.

\section{Experimental Results}
\label{sec:experiment}

%

Our architecture is evaluated on the CIFAR-10 dataset, which 
consists of \num{32}\texttimes\num{32} RGB images. The same 
preprocessing and data augmentation used in \cite{Resnet} is used 
for both training and testing. The model is trained for \num{400} 
epochs with a batch size of \num{256}, using the \ac{sgd} 
optimizer and cosine annealing as the learning rate scheduler.

Synthesizable C++ is used for both the hand-written layer process 
library and the Python-generated top-level dataflow code that 
calls the process functions. The implementation flow uses Xilinx 
Vitis HLS for \ac{rtl} code generation and Vivado for 
implementation on the Ultra96-v2 and Kria KV260 boards. 
\Cref{tab:resources} shows the available resources for the two 
boards.

\begin{table}
    \caption{Resources of the Ultra96 and Kria KV260 boards}
    \label{tab:resources}
    \centering
    \begin{adjustbox}{max width = 0.475\textwidth}
    \begin{tabular}{lccrrrrr}
        \toprule
        \multicolumn{1}{c}{\textbf{Board}}
            & \phantom{a}
                & \textbf{FPGA part}
                    & \textbf{LUT}
                        & \textbf{FF}
                            & \textbf{BRAM}
                                & \textbf{DSP}
                                    & \textbf{URAM} \\
        \midrule
        Ultra96 && xczu3eg & 141120 & 70560 & 216 & 360 & 0 \\
        Kria KV260 && xczu5eg & 234240 & 117120 & 144 & 1248 & 64 \\
        \bottomrule
    \end{tabular}
    \end{adjustbox}
\end{table}


The obtained throughputs (\acs{fps}, \si{\giga ops \per \second}) 
and the latency (\si{\milli\second}) are shown in 
\cref{tab:results_perf}.
The final resource utilization is summarized in 
\cref{tab:results_res}.

\begin{table*}
    \caption{Performance for CIFAR-10 on KV260 and Ultra96-v2.}
    \label{tab:results_perf}
    \centering
    \footnotesize
\begin{tabular}{
    l
    l
    c
    c
    c
    c
    c
    c
    c
}
    \toprule
    \multicolumn{1}{@{}c}{\textbf{Model}} &
        \multicolumn{1}{c}{\textbf{FPGA}} &
            \multicolumn{1}{c}{{Bit}} &
                \multicolumn{1}{c}{\textbf{Freq.}} &
                    \multicolumn{1}{c}{\textbf{Throughput}} &
                        \multicolumn{1}{c}{\textbf{Throughput}} &
                            \multicolumn{1}{c}{\textbf{Latency}} &
                                \multicolumn{1}{c}{\textbf{Power}} &
                                    \multicolumn{1}{c@{}}{\textbf{Acc.}} \\
    &
        &
            &
                \multicolumn{1}{c}{(\si{\mega\hertz})} &
                    \multicolumn{1}{c}{(\acs{fps})} &
                        \multicolumn{1}{c}{(\si{\giga ops\per\second})} &
                            \multicolumn{1}{c}{(\si{\milli\second})} &
                                \multicolumn{1}{c}{(\si{\watt})} &
                                    \multicolumn{1}{c@{}}{(\si{\percent})} \\
    \midrule
    ResNet20 CNN\textsuperscript{\dag} \cite{Zhang2022} &
        KV260 &
            8 &
                200 &
                    N/A &
                        214 &
                            1.221 &
                                1.07\textsuperscript{\dag} &
                                    90.8 \\
    AdderNet\textsuperscript{\dag} \cite{Zhang2022} &
        KV260 &
            8 &
                200 &
                    N/A &
                        317 &
                            0.624 &
                                1.52\textsuperscript{\dag} &
                                    89.9 \\
    ResNet20 CNN (our) &
        KV260  &
            8 &
                274 &
                    \textbf{7601} &
                        \textbf{616} &
                            \textbf{0.318} &
                                3.61 &
                                    91.3 \\
    \midrule
    ResNet8 CNN FINN \cite{framework2023} &
        KV260 &
            4 &
                225 &
                    13475 &
                        330 &
                            0.154 &
                                5.89 &
                                    85.9 \\
    ResNet8 CNN Vitis AI \cite{framework2023} &
        KV260 &
            8 &
                200 &
                    4458 &
                        109 &
                            1.293 &
                                6.42 &
                                    89.2 \\
    ResNet8 CNN (our) &
        KV260 &
            8 &
                274 &
                    \textbf{30153} &
                        \textbf{773} &
                            \textbf{0.046} &
                                3.60 &
                                    88.7 \\
    \midrule
    AdderNet \cite{manovel} &
        Ultra96 &
            16 &
                100 &
                    N/A &
                        N/A &
                            N/A &
                                N/A &
                                    91.3 \\
    ResNet20 CNN (our) &
        Ultra96 &
            8 &
                214 &
                    \textbf{3254} &
                        \textbf{264} &
                            \textbf{0.807} &
                                1.04 &
                                    \itshape 91.3 \\
    ResNet8 CNN (our) &
        Ultra96 &
            8 &
                214 &
                    \textbf{12971} &
                        \textbf{317} &
                            \textbf{0.111} &
                                0.56 &
                                    \itshape 88.7 \\
    \bottomrule
    \multicolumn{9}{l}{\scriptsize\textsuperscript{\dag} The description of how to measure power consumption is not explicitly provided, and the recorded idle consumption of the board does not align with the reported values.}
\end{tabular}\\
\end{table*}

\begin{table*}
    \caption{Resource utilization for CIFAR-10 on Kria KV260 and Ultra96-v2}
    \label{tab:results_res}
    \centering
    \begin{adjustbox}{max width=\linewidth}
        \begin{tabular}{
    l
    l
    S[table-format=1.0]
    S[table-format=2.1,input-symbols={()\%},table-space-text-post={\enspace(11.1\,\%)},table-align-text-post=true]
    S[table-format=2.1,input-symbols={()\%},table-space-text-post={\enspace(11.1\,\%)},table-align-text-post=true]
    S[table-format=2.1,input-symbols={()\%},table-space-text-post={\enspace(11.1\,\%)},table-align-text-post=true]
    S[table-format=3.0,input-symbols={()\%},table-space-text-post={\enspace(11.1\,\%)},table-align-text-post=true]
    S[table-format=3.1,input-symbols={()\%},table-space-text-post={\enspace(11.1\,\%)},table-align-text-post=true]
    S[table-format=2.0,input-symbols={()\%},table-space-text-post={\enspace(11.1\,\%)},table-align-text-post=true]
}
   \toprule
    \multicolumn{1}{c}{\textbf{Model}} &
        \multicolumn{1}{c}{\textbf{FPGA}} &
            \multicolumn{1}{c}{\textbf{Bit}} &
                \multicolumn{1}{c}{\textbf{kLUT}} &
                    \multicolumn{1}{c}{\textbf{kLUTRAM}} &
                        \multicolumn{1}{c}{\textbf{kFF}} &
                            \multicolumn{1}{c}{\textbf{DSP}} &
                                \multicolumn{1}{c}{\textbf{BRAM}} &
                                    \multicolumn{1}{c}{\textbf{URAM}} \\
    \midrule
    ResNet20 CNN \cite{Zhang2022} &
        KV260 &
            8 &
                41.8\hfill(\SI{35.7}{\percent}) &
                    17.6\hfill(\SI{30.1}{\percent}) &
                        34\hfill(\SI{14.5}{\percent}) &
                            545\hfill(\SI{43.7}{\percent}) &
                                40\hfill(\SI{27.7}{\percent}) &
                                    N/A \\
    AdderNet \cite{Zhang2022} &
        KV260 &
            8 &
                67.4\hfill(\SI{57.6}{\percent}) &
                    22.2\hfill(\SI{38.6}{\percent}) &
                        43.2\hfill(\SI{19.1}{\percent}) &
                            609\hfill(\SI{48.8}{\percent}) &
                                40\hfill(\SI{27.7}{\percent}) &
                                    N/A \\
    ResNet20 CNN (our) &
        KV260 &
            8 &
                81.2\hfill(\SI{69.4}{\percent}) &
                    11.8\hfill(\SI{20.5}{\percent}) &
                        83.5\hfill(\SI{35.6}{\percent}) &
                            626\hfill(\SI{50.2}{\percent}) &
                                73.5\hfill(\SI{51}{\percent}) &
                                    64\hfill(\SI{100}{\percent}) \\
    \midrule
    ResNet8 CNN-FINN \cite{framework2023} &
        KV260 &
            4 &
                81.4\hfill(\SI{69.5}{\percent}) &
                    N/A &
                        87.6\hfill(\SI{37.4}{\percent}) &
                            N/A &
                                28.5\hfill(\SI{19.8}{\percent}) &
                                    N/A \\
    ResNet8 CNN-VITIS AI \cite{framework2023} &
        KV260 &
            8 &
                25.6\hfill(\SI{21.8}{\percent}) &
                    N/A &
                        33.7\hfill(\SI{14.4}{\percent}) &
                            110\hfill(\SI{8.8}{\percent}) &
                                8.8\hfill(\SI{8.8}{\percent}) &
                                    18\hfill(\SI{28.1}{\percent}) 
    \\
    ResNet8 CNN (our) &
        KV260 &
            8 &
                74.6\hfill(\SI{63.7}{\percent}) &
                    8.7\hfill(\SI{15.1}{\percent}) &
                        75.7\hfill(\SI{32.3}{\percent}) &
                            773\hfill(\SI{61.9}{\percent}) &
                                98.0\hfill(\SI{68.1}{\percent}) &
                                    63\hfill(\SI{98.4}{\percent}) \\
    \midrule
    AdderNet\cite{manovel} &
        Ultra96 &
            16 &
                66.6\hfill(\SI{95.9}{\percent}) &
                    N/A &
                        17.8\hfill(\SI{2.9}{\percent}) &
                            0\hfill(\SI{0}{\percent}) &
                                N/A &
                                    N/A \\
    ResNet20 CNN (our) &
        Ultra96 &
            8 &
                54.4\hfill(\SI{77.1}{\percent}) &
                    10.2\hfill(\SI{35.6}{\percent}) &
                        57.6\hfill(\SI{40.8}{\percent}) &
                            318\hfill(\SI{88.3}{\percent}) &
                                89.5\hfill(\SI{41.4}{\percent}) &
                                    0\hfill(\SI{0}{\percent}) \\
    ResNet8 CNN (our) &
        Ultra96 &
            8 &
                46.4\hfill(\SI{65.8}{\percent}) &
                    6.2\hfill(\SI{21.5}{\percent}) &
                        45.1\hfill(\SI{32.0}{\percent}) &
                            360\hfill(\SI{100}{\percent}) &
                                54\hfill(\SI{25}{\percent}) &
                                    0\hfill(\SI{0}{\percent}) \\
    \bottomrule
\end{tabular}
    \end{adjustbox}
\end{table*}

Our proposed architecture is first compared with a ResNet20 
implementation and the derived AdderNet described in \cite{Zhang2022}, which is are 
the most efficient \ac{cnn} implementations on \acp{fpga} in terms \ac{dsp} packing and 
model architecture to-date.
Our implementation achieves speedups (\si{\giga ops \per \second}) of 
\num{2.88}\texttimes\ and \num{1.94}\texttimes\ with 
\SI{0.5}{\percent} and \SI{1.4}{\percent} higher accuracy, with respect to the ResNet20 and Addernet in \cite{Zhang2022}, using the Kria KV260 as a reference platform.
Also, the latency is reduced by \num{3.84}\texttimes\ and \num{1.96}\texttimes\ respectively.

We then compare our results with the implementations of the 
ResNet8 model by Vitis AI and FINN described in \cite{framework2023}. Our 
solution achieves speedups of \num{6.8}\texttimes\ and 
\num{2.2}\texttimes\ with a latency improvement of \num{28.1}\texttimes\ and \num{3.35}\texttimes\, respectively. 
Vitis AI achieved better 
accuracy by \SI{0.5}{\percent}, probably because it executes \emph{batch normalization} in hardware, while our implementation 
outperformed a 4-bit FINN implementation by \SI{2.8}{\percent}.

\section{Conclusion}
\label{sec:conc}


This work presents a design flow for \acp{cnn} specifically 
optimized for residual networks. It supports the most commonly 
used operations for classic \acp{cnn}, including convolutions, 
fully connected (linear) layers, batch normalization, ReLU 
activation functions, max/average pooling, and skip connections. 
It is also fairly platform-independent,  since it is based on 
heavily templatized layer models and comes with an \ac{ilp}-based 
optimization method to maximize throughput under resource 
constraints. This allows it to be used with various \ac{fpga} 
platforms, including embedded ones. 

A dataflow pipelined architecture minimizes buffering resources 
for networks with skip connections. The design is validated by 
experiments on ResNet8 and ResNet20 using the CIFAR-10 dataset. 
Both activations and weights are quantized in INT8 format using 
power-of-two scaling factors.

The design uses PyTorch and Brevitas for training and 
quantization, and Vitis HLS and Vivado for hardware 
implementation on Kria KV-260 and Ultra96-v2 boards.

The solution achieves an accuracy of \SI{88.7}{\percent} for 
ResNet8 and \SI{91.3}{\percent} for ResNet20, with throughputs of 
\num{12971} \acs{fps} and \num{3254} \acs{fps} on the Ultra96, and 
\num{30153} \acs{fps} and \num{7601} \acs{fps} on the Kria KV260. 
Compared to the state-of-the-art for \acs{cnn} residual network 
acceleration on \acp{fpga} \cite{Zhang2022}, it achieves 
\num{2.88}\texttimes\ speedup with \SI{0.5}{\percent} higher 
accuracy for ResNet20 and \num{2.2}\texttimes\ speedup with 
\SI{2.8}{\percent} higher accuracy for ResNet8 
\cite{framework2023}. Compared to a residual network with packed 
adders \cite{Zhang2022}, it achieves \num{1.94}\texttimes\ 
speedup with \SI{1.4}{\percent} higher accuracy and a latency improvement of \SI{1.96}{\times}. 
Considering state-of-the-art frameworks, the comparison shows that the 
resource-efficient implementation of the residual layer achieves 
a Pareto-optimal implementation for accuracy, throughput, and latency. Since the boards are the same and all approaches utilize most resources of each \ac{fpga}, lower latency also means lower energy than the state of the art.

In summary, the proposed design architecture shows potential as 
an alternative to the commonly used residual network accelerators 
on platforms with limited resources. It delivers greater 
throughput and energy efficiency than the state-of-the-art 
without increasing hardware costs.

\newpage

\bibliographystyle{abbrv}
\bibliography{refs}
\end{document}